\documentclass[12pt,reqno]{amsart}
\usepackage{graphicx}
\usepackage{amssymb,amsmath}
\usepackage{amsthm}
\usepackage{color}
\usepackage[pdf]{pstricks}
\usepackage{hyperref}

\usepackage{amssymb}
\usepackage{epsfig}
\usepackage{extarrows}
\usepackage{amsfonts}
\usepackage{graphicx}

\newtheorem{theorem}{Theorem}

\begin{document}
	
\title[BO breathers]{\bf Bright and dark breathers of the Benjamin--Ono equation on the traveling periodic background}
	
\author{Jinbing Chen}
\address[J. Chen]{School of Mathematics, Southeast University, Nanjing, Jiangsu 210096, PR China}
\email{cjb@seu.edu.cn}

\author{Dmitry E. Pelinovsky}
\address[D. Pelinovsky]{Department of Mathematics and Statistics, McMaster University, Hamilton, Ontario, Canada, L8S 4K1}
\email{dmpeli@math.mcmaster.ca}

\begin{abstract}
The Benjamin--Ono (BO) equation describes long internal waves of small amplitude in deep fluids. Compared to its counterpart for shallow fluids, the Korteweg-de Vries (KdV) equation, the BO equation admits exact solutions for the traveling periodic and solitary waves as well as their interactions expressed in 
elementary (trigonometric and polynomial) functions. Motivated by a recent progress for the KdV equation, we discover here two scenarios of the soliton-periodic wave interactions which result in the propagation of either elevation (bright) or depression (dark) breathers (periodic in time coherent structures). The existence of two different breathers is related to the band-gap spectrum of the Lax operator associated with the traveling periodic wave. Given a simple structure of the exact solutions in the BO equation, we obtain a closed-form expression for multi-solitons interacting with the traveling periodic wave. 
\end{abstract}

\date{\today}
\maketitle

{\em This article is written in memory of N. F. Smyth for his many contributions to studies of the modulation theory and, in particular, periodic waves of the Benjamin--Ono equation.}

\section{Introduction}

Two limiting cases of long internal waves of small amplitude can be described within the integrable evolution equations: the Korteweg--de Vries (KdV) equation for shallow fluids and the Benjamin--Ono (BO) equation for deep fluids \cite{Grimshaw}. These integrable equations have a large class of exact solutions describing spatially periodic and solitary traveling waves as well as their interactions \cite{Matsuno-book}. 

Exact solutions for the time-periodic interactions of a single solitary wave with the spatially periodic traveling wave were constructed recently for the KdV equation in \cite{Bertola,HMP} after much earlier work \cite{Gesztesy,Kuznetsov}. Such periodic in time coherent structures were referred to as breathers. Two families of breathers were constructed as elevation (bright) breathers and  depression (dark) breathers. The two families correspond to two scenarios of a more complicated  interaction of a solitary wave with dispersive shock waves (DSWs), see a recent review in \cite{Cole}. Bright breathers describe a transmission of a solitary wave overtaking DSWs and dark breathers describe a trapping of a solitary wave in DSWs. The exact bright and dark breather solutions of the KdV equation expressed in terms of the Jacobi theta functions were used for comparison with numerical and laboratory experiments of soliton-DSW interactions in shallow fluids \cite{Maiden,Mao,Sprenger}. 

Similar breathers of the defocusing nonlinear Schr\"{o}dinger equation were obtained in different solution forms in \cite{Ling,Lou,Shin,Takahashi}. Recently, breathers of the defocusing modified KdV equation were elaborated in \cite{HMP-24} by analyzing properties of the Jacobi theta functions. 
 
The purpose of this work is to obtain breather solutions for  the BO equation which covers applications of soliton-DSW interactions in deep fluids. See \cite{Trillo13,Smyth02} for two particular examples where the DSWs of the BO equation arise in various physical applications.

We take the BO equation in the standard normalized form: 
\begin{equation}
\label{BO}
u_t + 2 u u_x + H(u_{xx}) = 0, \quad H(f) := \frac{1}{\pi} {\rm p. v.} \int_{-\infty}^{\infty} \frac{f(y) dy}{y-x}.
\end{equation}
The periodic traveling wave is expressed in terms of elementary functions:
\begin{equation}
\label{per-wave}
u(x,t) = \frac{k \sinh \phi}{\cos(k\xi) + \cosh \phi},
\end{equation}
where $\xi = x - c_0 t - \xi_0$ is the traveling wave coordinate, $k, \phi > 0$ are arbitrary parameters, $\xi_0 \in \mathbb{R}$ is an arbitrary phase, and $c_0 = k \coth \phi$. The solitary wave corresponds to the limit $\phi \to 0$ with $k = c_0 \tanh \phi \to 0$ uniquely determined by fixed $c_0 > 0$. After phase transformation $\xi \to \xi + \pi k^{-1}$, the periodic wave solution (\ref{per-wave}) reduces to the solitary wave solution written in the form:
\begin{equation}
\label{sol-wave}
u(x,t) = \frac{2c_0}{1 + c_0^2 \xi^2}.
\end{equation}

The main result of this work is that the BO equation admits two families of breather solutions similar to the KdV equation. However, the breather solutions of the BO equation are expressed in terms of elementary (trigonometric and power) functions instead of the Jacobi theta functions. The new exact solutions are given by 
\begin{equation}
\label{breather}
\hat{u}(x,t) = \frac{2 (c + k \beta) \cosh \phi + [k (1+ \beta^2 + c^2 \eta^2) + 2 \beta c] \sinh \phi + 2 c \cos(k\xi)}{(1 + \beta^2 + c^2 \eta^2) \cosh \phi + 2 \beta \sinh \phi + (1 - \beta^2 + c^2 \eta^2) \cos(k\xi)  + 2 \beta c \eta \sin(k\xi)},
\end{equation}
where $\eta = x - c t - \eta_0$ is the traveling wave coordinate, 
$\eta_0 \in \mathbb{R}$ is an arbitrary phase, $\beta$ is uniquely defined by 
\begin{equation}
\label{b-definition}
\beta = \frac{2ck}{(c-c_0)^2 - k^2},
\end{equation}
and $c > 0$ is an arbitrary speed of the solitary wave defined in two disjoint regions: 
\begin{equation}
\label{regions}
(\mbox{\rm BB}) \;\; c - c_0 > k \quad \mbox{\rm and} \quad (
\mbox{\rm DB}) \;\; c-c_0 < -k.
\end{equation}
The bright breathers (BB) correspond to $c > c_0 + k$ and the dark breathers (DB) correspond to $0 < c < c_0 - k$, hence the former overtakes the periodic wave and the latter is trapped in the periodic wave, similarly to the KdV equation \cite{HMP}. As $\eta \to \pm \infty$, $\hat{u}(x,t) \to u(x,t)$ given by (\ref{per-wave}) so that no phase shift arises in the interactions between the solitary wave and the periodic wave, contrary to the KdV equation \cite{HMP}. The localization of the 
solitary wave is defined by its speed $c$: the larger $c$ is, the smaller is the width of the solitary wave. 

\begin{figure}[htb!]
	\centering
	\includegraphics[width=7.5cm,height=5cm]{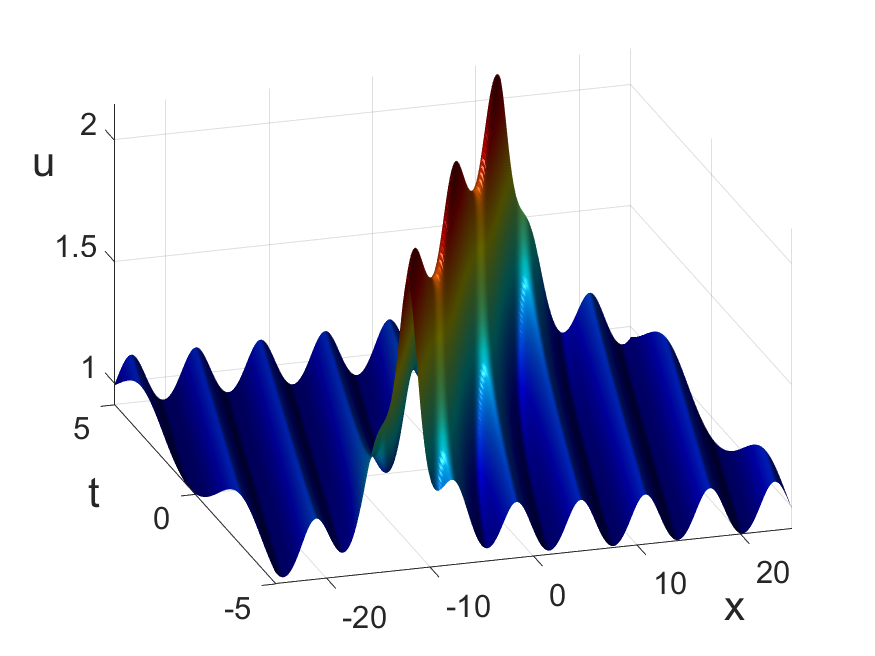}
	\includegraphics[width=7.5cm,height=5cm]{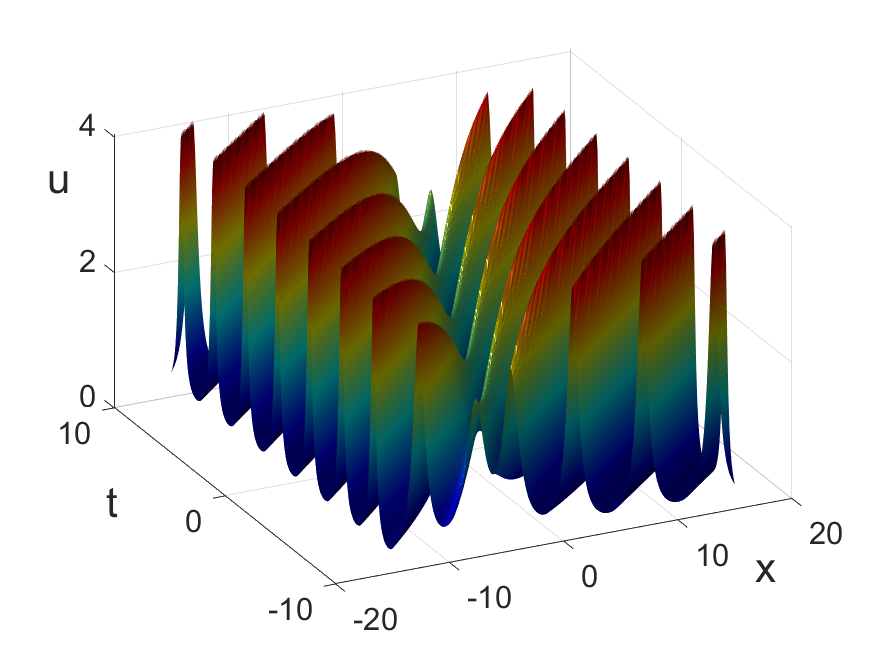}
	\caption{Left: bright breather with $k = 1$, $\phi = 3$, and $c = c_0 + 1.5k$. Right: dark breather with $k = 1$, $\phi = 0.5$, and $c = c_0 - 1.5k$.}
	\label{fig-sol-surface}
\end{figure}

Figure \ref{fig-sol-surface} show the solution surfaces computed from the exact solution (\ref{breather}) with two choices of $c$ in (\ref{regions}). 
The left panel shows propagation of a bright (elevation) breather with the speed exceeding the speed of the background periodic wave. The right panel shows propagation of a dark (depression) breather with the speed being smaller that the speed of the background periodic wave. 

By generalizing the breather solution (\ref{breather}) in the determinant form \cite{Matsuno04}, we present the general multi-breather solution on the 
traveling periodic wave (\ref{per-wave}) as the following theorem.

\begin{theorem}
	\label{theorem-main}
	The BO equation (\ref{BO}) admits the following general multi-breather solution on the traveling periodic wave (\ref{per-wave}) in the form 
\begin{equation}
\label{Hirota-intro}
u(x,t) = i \partial_x \log \frac{\det(F')}{\det(F)},  
\end{equation}
with 
$$
F = \left[ 
\begin{matrix} \displaystyle 
1 + e^{ik\xi + \phi} & \displaystyle \frac{2 k}{k+c_0 - c_1} & \dots & \displaystyle \frac{2 k}{k+c_0 - c_N} \\ \displaystyle
\frac{2 c_1}{k+c_1 - c_0} & \displaystyle -i c_1 \eta_1 - 1 - \frac{2k c_1}{(c_0-c_1)^2-k^2} & \dots & \displaystyle \frac{2 c_1}{c_1 - c_N} \\
\vdots & \vdots & \ddots & \vdots \\ \displaystyle
\frac{2 c_N}{k+c_N - c_0} & \displaystyle \frac{2 c_N}{c_N - c_1} & \dots & \displaystyle -i c_N \eta_N - 1 - \frac{2k c_N}{(c_0-c_N)^2-k^2}
\end{matrix}
\right]
$$
and
$$
F' = \left[ 
\begin{matrix} \displaystyle 
1 + e^{ik\xi - \phi} & \displaystyle \frac{2 k}{k+c_0 - c_1} & \dots & \displaystyle \frac{2 k}{k+c_0 - c_N} \\ \displaystyle
\frac{2c_1}{k+c_1 - c_0} & \displaystyle -i c_1 \eta_1 + 1 - \frac{2k c_1}{(c_0-c_1)^2-k^2} & \dots & \displaystyle \frac{2 c_1}{c_1 - c_N} \\
\vdots & \vdots & \ddots & \vdots \\ \displaystyle
\frac{2 c_N}{k+c_N - c_0} & \displaystyle \frac{2 c_N}{c_N - c_1} & \dots & \displaystyle -i c_N \eta_N + 1 - \frac{2k c_N}{(c_0-c_N)^2-k^2}
\end{matrix}
\right]
$$
where for $1 \leq j \leq N$, we have defined $\eta_j = x - c_j t - x_j$ with arbitrary $x_j \in \mathbb{R}$ and arbitrary distinct $c_j > 0$ satisfing $|c_j - c_0| > k$. The solution (\ref{Hirota-intro}) is real-valued and bounded for every $(x,t) \in \mathbb{R}\times \mathbb{R}$.
\end{theorem}

\begin{figure}[htb!]
	\centering
	\includegraphics[width=5cm,height=4cm]{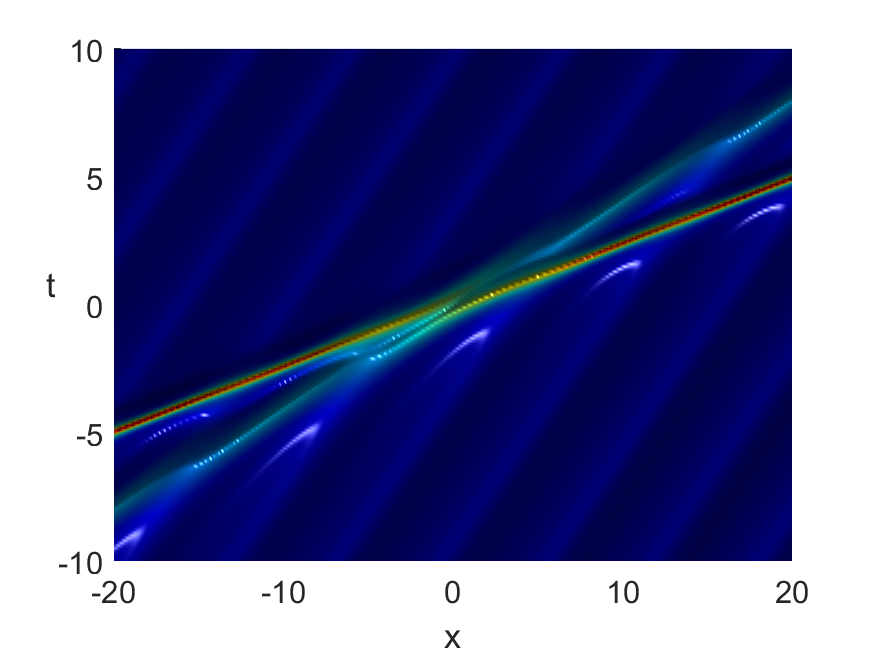}
	\includegraphics[width=5cm,height=4cm]{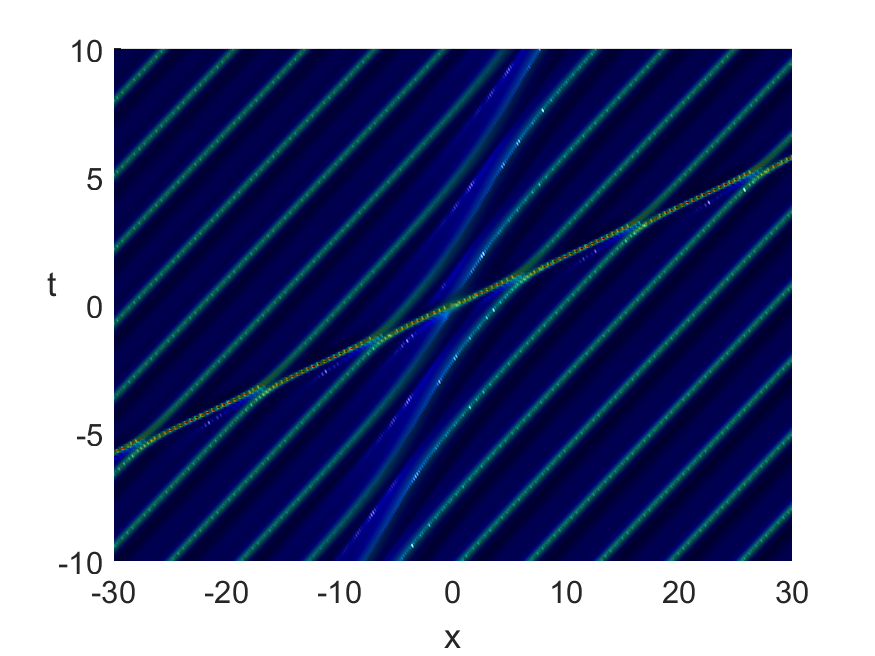}
	\includegraphics[width=5cm,height=4cm]{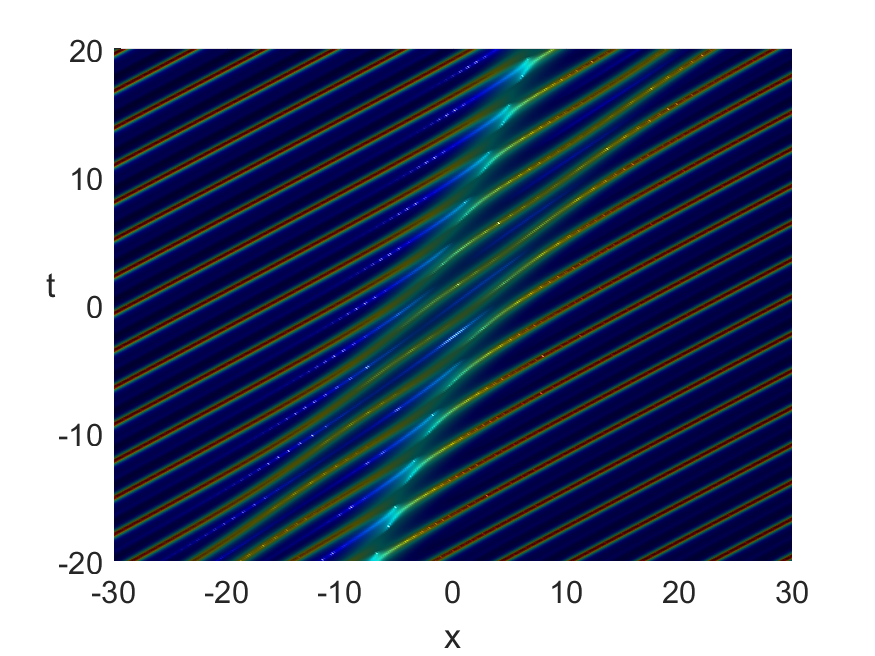}
	\caption{Left: bright--bright breather with $k = 1$, $\phi = 3$, $c_1 = c_0 + 1.5k$, and $c_2 = c_0 + 3k$. Middle: bright--dark breather with $k = 1$, $\phi = 0.5$, $c_1 = c_0 - 1.5 k$, and $c_2 = c_0 + 3k$. Right: dark--dark breather with $k = 1$, $\phi = 0.5$, $c_1 = c_0 - 1.75k$, and $c_2 = c_0 - 1.25k$.}
	\label{fig-two-solitons}
\end{figure}

The solution in Theorem \ref{theorem-main} corresponds to $N$ solitary waves propagating on the traveling periodic wave (\ref{per-wave}) with the speeds $c_1,\dots, c_N$. For $N = 2$, Figure \ref{fig-two-solitons} shows the top views of three solution surfaces computed from the exact solution (\ref{Hirota-intro}) with $N = 2$ for three choices of $c_1$ and $c_2$. Phase parameters $\xi_0$, $x_1$, and $x_2$ are all set to zero. The left panel shows propagation of two bright breathers for $c_1,c_2 > c_0 + k$. The middle panel shows propagation of one bright and one dark breather for $c_1 < c_0-k$ and $c_2 > c_0 + k$. The right panel shows propagation of two dark breathers for $c_1,c_2 < c_0 - k$.

Our work relies on several recent advances in the literature devoted to the periodic and solitary wave solutions of the BO equation (\ref{BO}).\\

\begin{enumerate}
	\item {\bf Dispersive shock waves (DSWs)} in the BO equation were studied with the Whitham modulation theory in \cite{DK-91,Smyth99,Matsuno98,MATSUNO}. The Whitham equations for the BO equation reduce to the Hopf (inviscid Burgers) equations. The breather solutions equivalent to the solution form (\ref{breather}) were obtained in Appendix B of \cite{Smyth99} but it was missed that two families of bright and dark breathers coexist on the background of the traveling periodic wave. These earlier results have motivated many recent works \cite{Smyth18,GSS21,MSKK,Smyth21}, where the DSWs have been studied in the nonlocal generalizations of the BO equation. \\	
	
	\item {\bf Lax spectrum} associated with traveling periodic waves was rigorously justified  only recently in \cite{GK-21} and was generalized to other periodic potentials \cite{GKT-20,GKT-22}. In our notations, the Lax spectrum of the traveling periodic wave (\ref{per-wave})  is located at $[\lambda_0,\mu_0] \cup [0,\infty)$, where
	\begin{equation}
	\label{lax-spectrum}
\lambda_0 = -\frac{1}{2} (c_0 + k), \qquad \mu_0 = -\frac{1}{2} (c_0 - k).
	\end{equation}
The two families of the bright (dark) breathers (\ref{regions}) are associated 
with the semi-infinite (finite) spectral gap in the Lax spectrum: 
\begin{equation}
\label{DT-outcome}
(\mbox{\rm BB}) \;\;-\frac{1}{2} c \in (-\infty,\lambda_0) \quad \mbox{\rm and} \quad (
\mbox{\rm DB}) \;\; -\frac{1}{2} c \in (\mu_0,0),
\end{equation}
where $\lambda = -\frac{1}{2}c$ is the eigenvalue of the Lax spectrum associated with the single solitary wave  \cite{AF83,KM98}. The division of breathers into two families related to the choice of eigenvalue $\lambda$ in the semi-infinite versus finite spectral gaps of the Lax spectrum is very similar to the KdV equation \cite{HMP}. \\

\item {\bf Modulational stability} of traveling periodic waves of the BO equation was studied with the explicit computations of eigenfunctions of the linearized BO equation \cite{MS94} and the asymptotic reduction of the BO equation to the nonlocal defocusing NLS equation for long perturbations of the small-amplitude waves \cite{P95}. Modulational stability was later confirmed for the BO equation with a rigorous analysis of the spectral stability problem \cite{BH14}. Breather solutions like the one given by (\ref{breather}) may only propagate on the modulationally stable traveling periodic wave background in agreement with these stability results.  \\

\item {\bf Particle dynamics} of the Calogero--Moser system has been studied in relation to poles of the meromorphic solutions of the BO equation \cite{Case79}. In this context and in the presence of a harmonic potential, an interaction of a single solitary wave with $N$ solitary waves for large $N$ was studied in \cite{Abanov} (see their Figure 1). Interactions of solitary waves via the particle dynamics were also considered for the BO equation and its modifications in \cite{Abanov1,Abanov2}. Poles of the exact solution (\ref{breather}) describe the exact trajectories of particles of the Calogero--Moser system for such interactions in the BO equation (\ref{BO}). \\
\end{enumerate}

We note that the KdV equation and the BO equation do not admit conventional breather solutions which are localized in space and periodic in time. The 
proof of non-existence of breathers in these equations was recently given 
in \cite{Munoz1,Munoz2}. The breather wave packets derived here and in 
\cite{Bertola,HMP} are obtained as a result of periodic interaction 
of solitary waves and the traveling periodic wave. Both the KdV and BO equations are the limiting cases of the integrable intermediate long-wave (ILW) equation \cite{Miloh}. Similar breathers are expected to exist in the ILW equation but 
expressed in the Jacobi theta functions. Only the BO equation describes such complicated wave interactions in elementary  functions. 

The rest of this paper is organized as follows. The derivation of the single-breather solutions (\ref{breather}) is given in Section \ref{sec-2}. The relation between the existence region (\ref{regions})  and the Lax spectrum of the traveling periodic wave is explained in Section \ref{sec-3}. The exact solution (\ref{Hirota-intro}) for 
multi-breather solutions is derived in  Section \ref{sec-4}. 

\section{Derivation of the single-breather solutions}
\label{sec-2}

In order to obtain the breather solutions (\ref{breather}), we follow the multi-periodic solutions obtained in the bilinear method \cite{Satsuma79}. Equivalently the same solutions can be obtained by the B\"{a}cklund--Darboux transformation of the BO equation \cite{Nakamura79} with an eigenvalue $\lambda = -\frac{1}{2} c$ in either of the two intervals (\ref{DT-outcome}). 

The class of multi-periodic and multi-soliton solutions of the BO equation can be obtained in the form:
\begin{equation}
\label{Hirota}
u(x,t) = i \partial_x \log \frac{f'(x,t)}{f(x,t)}, 
\end{equation}
where $f$ ($f'$) has zeros only in the upper (lower) half-plane of $z := x + iy$ and satisfies the bilinear equation 
\begin{equation}
\label{Hirota-eq}
i (f'_t f - f' f_t) = f'_{xx} f - 2 f'_x f_x + f' f_{xx}.
\end{equation}

Let us now consider the simplest solutions of the bilinear equation (\ref{Hirota-eq}). \\

{\bf 1-periodic solutions.} The traveling periodic wave (\ref{per-wave}) is obtained from (\ref{Hirota}) by using the following solution of the bilinear equation (\ref{Hirota-eq}):
\begin{equation}
\label{1-periodic}
\left\{ \begin{array}{l} f(x,t) = 1 + e^{ik \xi + \phi}, \\
f'(x,t) = 1 + e^{ik \xi - \phi}, \end{array} \right.
\end{equation}
where $\xi = x - c_0t - \xi_0$,  $k, \phi > 0$, $\xi_0 \in \mathbb{R}$, and $c_0 = k \coth \phi > 0$. \\

{\bf 2-periodic solutions.} The $2$-periodic solutions are generated from the following solution  of the bilinear equation (\ref{Hirota-eq}):
\begin{equation}
\label{2-periodic}
\left\{ \begin{array}{l} f(x,t) = 1 + \alpha e^{ik \xi + \phi} + \alpha e^{i \kappa \eta + \psi} + e^{i k \xi + i \kappa \eta + \phi + \psi}, \\
f'(x,t) = 1 +  \alpha e^{ik \xi - \phi} + \alpha e^{i \kappa \eta - \psi} + e^{i k \xi + i \kappa \eta - \phi - \psi}, \end{array} \right.
\end{equation}
where $\xi = x - c_0t - \xi_0$, $\eta = x - c t - \eta_0$,  $k \phi > 0$,  $\kappa \psi > 0$,  $\xi_0, \eta_0 \in \mathbb{R}$, $c_0 = k \coth \phi > 0$, $c = \kappa \coth \psi > 0$, and 
$$
\alpha = \frac{\sqrt{(c-c_0)^2 - (\kappa + k)^2}}{\sqrt{(c-c_0)^2 - (\kappa - k)^2}}.
$$
Since 
$$
f'(x,t) = e^{i k \xi + i \kappa \eta - \phi - \psi} \bar{f}(x,t),
$$
with bar being the complex conjugation, the transformation (\ref{Hirota}) with $f$, $f'$ in (\ref{2-periodic}) generates a real solution $u(x,t)$ of the BO equation (\ref{BO}). Moreover, it was shown in \cite{Satsuma79} that 
the solution $u(x,t)$ is bounded for every $(x,t) \in \mathbb{R}\times \mathbb{R}$ and the assumption on zeros of $f$ ($f'$) in the upper (lower) half-plane is satisfied if parameters $c$, $c_0$, $\kappa$, and $k$ satisfy the constraint 
\begin{equation}
\label{constraint}
(c - c_0)^2 > (|k| + |\kappa|)^2.
\end{equation}
\vspace{0.1cm}

{\bf Breather solutions.} In the limit $\kappa \to 0$, there are two ways to satisfy the constraint (\ref{constraint}): $c - c_0 > k$ or $c - c_0 < -k$ with $k > 0$ assumed here. This choice leads to the two disjoint solution families (\ref{regions}). The breather solution (\ref{breather}) arises in the limit $\psi \to 0$ with $\kappa = c \tanh \psi \to 0$ uniquely defined for fixed $c > 0$. Expanding $\alpha$ yields 
\begin{align*}
\alpha &= 1 - \frac{2 k \kappa}{(c-c_0)^2 - k^2} + \frac{2 k^2 \kappa^2}{[(c-c_0)^2 - k^2]^2} + \mathcal{O}(\kappa^3) \\
&= 1 - \beta \psi + \frac{1}{2} \beta^2 \psi^2 + \mathcal{O}(\psi^3),  
\end{align*}
where $\beta > 0$ is given by (\ref{b-definition}). Using the transformation $\eta \to \eta + \pi \kappa^{-1}$ and expanding as $\psi \to 0$ yield at the leading order:
\begin{equation}
\label{expansion-for-1}
\left\{ \begin{array}{l} 
\displaystyle 
f(x,t) = \beta \psi (1 -  e^{ik \xi + \phi}) - \psi (1 + i c \eta) (1 + e^{ik \xi + \phi}) + \mathcal{O}(\psi^2), \\
\displaystyle 
f'(x,t) =  \beta \psi (1 -  e^{ik \xi - \phi}) + \psi (1 - i c \eta) (1 + e^{ik \xi - \phi}) + \mathcal{O}(\psi^2). \end{array} \right.
\end{equation}
Using (\ref{Hirota}) with the leading $\mathcal{O}(\psi)$ order of $f$ and $f'$ yields the breather solutions (\ref{breather}) which are obviously real. To show that the breather solutions (\ref{breather}) are bounded for every $(x,t) \in \mathbb{R}\times \mathbb{R}$ and the assumption on zeros of $f$ ($f'$) in the upper (lower) half-plane is satisfied, we consider the zeros of $f(x,t)$ at the leading $\mathcal{O}(\psi)$ order in (\ref{expansion-for-1}), that is, 
$$
\beta (1 -  e^{ik \xi + \phi}) -  (1 + i c \eta) (1 + e^{ik \xi + \phi}) = 0.
$$
Zeros are defined from 
$$
e^{i k \xi + \phi} = \frac{\beta - 1 - i c \eta}{\beta + 1 + i c \eta}
$$
for which 
$$
{\rm Im}(\xi) = k^{-1} (\phi + q), \quad q := \frac{1}{2} \log \frac{(\beta + 1)^2 + c^2 \eta^2}{(\beta - 1)^2 + c^2 \eta^2}.
$$
Since $k, \beta, \phi, q > 0$, zeros of $f$ are located in the upper half-plane away from the real axis. \\

{\bf $2$-soliton solutions.} The $2$-soliton solutions arise from the breather solutions in the limit $\phi \to 0$ 
with $k = c_0 \tanh \phi \to 0$ uniquely defined for fixed $c_0 > 0$. After the transformation $\xi \to \xi + \pi k^{-1}$, the expansion (\ref{expansion-for-1}) is further reduced to the form:
\begin{equation*}
\left\{ \begin{array}{l} 
\displaystyle f(x,t) = \left[ 
(1+ic_0 \xi) (1+ic \eta) + \frac{4c c_0}{(c-c_0)^2} \right] \phi \psi + \mathcal{O}(\phi^2 \psi^2), \\
\displaystyle 
f'(x,t) =  \left[ (1-ic_0 \xi) (1-ic \eta) + \frac{4c c_0}{(c-c_0)^2} \right] \phi \psi + \mathcal{O}(\phi^2 \psi^2). \end{array} \right.
\end{equation*}
The $2$-soliton solutions are well-known from  \cite{Matsuno82,Nakamura79,Satsuma79}. After the limiting transition $k \to 0$ in (\ref{regions}), the bright breathers correspond to $c > c_0$ and the dark breathers correspond to $0 < c < c_0$, where $c_0$ and $c$ are wave speeds of the two distinct solitary waves. This agrees with the property that the wave amplitudes are proportional to the wave speeds. The new soliton's speed $c$ exceeds the background soliton's speed $c_0$ for a bright breather which degenerates into a larger soliton overtaking the background soliton. The new soliton's speed $c$ is smaller than the background soliton's speed $c_0$ for a dark breather which degenerates into a smaller soliton staying behind the background soliton. 

\section{Existence of breathers and the Lax spectrum of the periodic wave}
\label{sec-3}

The BO equation (\ref{BO}) arises as a compatibility condition 
of the following system of Lax equations \cite{AF83,KM98}:
\begin{equation}
\label{Lax-pair}
\left\{ \begin{array}{l} i \varphi^+_x + \lambda( \varphi^+ - \varphi^-) + u \varphi^+ = 0, \\
i \varphi^{\pm}_t - 2 i \lambda \varphi^{\pm}_x + \varphi^{\pm}_{xx} - [\pm i u_x + H(u_x)] \varphi^{\pm} = 0,
\end{array} \right.
\end{equation}
where $\varphi^{\pm}$ are analytic functions in $\mathbb{C}^{\pm}$. If $u(x+L) = u(x)$ is periodic in $x$ with period $L = \frac{2\pi}{k}$ as in (\ref{per-wave}), then Lax spectrum can be defined from admissible values of the linear operator acting on $\varphi^+$:
\begin{equation}
\label{Lax-operator}
L_u := -i \partial_x - P^+(u \; \cdot), 
\end{equation}
where $P^+ = \frac{1}{2} (1 - i H)$ is the projection operator acting on functions defined on the real line and returning functions which can be analytically continued in $\mathbb{C}^+$. 

As is shown in Proposition 2.2 of \cite{GK-21}, the Lax spectrum of $L_u$ on the periodic domain consists of simple eigenvalues enumerated as $\{ \lambda_n \}_{n = 0}^{\infty}$ such that $\lambda_{n+1} \geq \lambda_n + k$ for $n \geq 0$, where $k = \frac{2\pi}{L}$ ($k = 1$ was used in \cite{GK-21} with $L = 2\pi$). On the other hand, as is shown in Proposition C.2 of \cite{GK-21}, the Lax spectrum of $L_u$ on the real line is absolutely continuous and consists of a union of bands 
$\cup_{n=0}^{\infty} [\lambda_n, \lambda_n + k]$.

The traveling periodic wave (\ref{per-wave}) corresponds to a particular case of the one-gap potential for which $\lambda_{n+1} = \lambda_n + k$ for every $n \in \mathbb{N}$. As follows from the computations in Appendix B of \cite{GK-21} 
(generalized in our notations), that $\lambda_0$ is given by (\ref{lax-spectrum}) and $\lambda_n = n-1$ for $n \in \mathbb{N}$. Then, 
$\mu_0 = \lambda_0 + k$ is defined in (\ref{lax-spectrum}) in agreement with formula (1.19) in \cite{DK-91}, as well as Proposition C.2 in \cite{GK-21}.

\vspace{0.4cm}
\centerline{\fbox{\parbox[cs]{1,0\textwidth}{
		To summarize, the Lax spectrum for the traveling periodic wave (\ref{per-wave}) is located in $[\lambda_0,\mu_0] \cup [0,\infty)$. The existence region (\ref{regions}) of the breather solutions (\ref{breather}) corresponds to the two choices in (\ref{DT-outcome}) for $\lambda = -\frac{1}{2}c$, the eigenvalue of the point spectrum associated to the breather.
		}}}
\vspace{0.4cm}

The rest of this section contains computations of the explicit expressions for eigenfunctions of the Lax system (\ref{Lax-pair}), which visualize the general results above. As an interesting new feature, we discover the following properties:\\
\begin{enumerate}
	\item[(P1)] Eigenvalues $\{ \lambda_n \}_{n = 0}^{\infty}$ are defined from the zero-mean condition on $\varphi^-$ among the $L$-periodic and $\mathbb{C}^{\pm}$-analytic eigenfunctions $\varphi^{\pm}$.\\
	
	\item[(P2)] Eigenvalues $\{ \mu_n \}_{n=0}^{\infty}$ with $\mu_n = \lambda_n + k$ is defined from the zero-mean condition on $\varphi^+$ among the $L$-periodic and $\mathbb{C}^{\pm}$-analytic eigenfunctions $\varphi^{\pm}$. \\
	
	\item[(P3)] The $L$-periodic and $\mathbb{C}^{\pm}$-analytic eigenfunctions $\varphi^{\pm}$ without the zero-mean conditions exist for every $\lambda \in \mathbb{R}$.\\
\end{enumerate}

There exist two general solutions of the linear system (\ref{Lax-pair}). The first solution has $\varphi^- \equiv 0$ and the second solution has both components $\varphi^+$ and $\varphi^-$ nonzero. \\

If $\varphi^- \equiv 0$, the exact solution of the linear system (\ref{Lax-pair}) exists in the form:
\begin{equation}
\label{eigenfunction-1}
\varphi^+(x,t) = e^{i \lambda x + i \lambda^2 t} \frac{1 + e^{i k \xi + \phi}}{1 + e^{ik\xi - \phi}} =: e^{i \lambda x + i \lambda^2 t} p^+(\xi,t),
\end{equation}
where $p^+(\cdot,t)$ is an $L$-periodic and $\mathbb{C}^+$-analytic function with period $L = 2\pi/k$ for every $t \in \mathbb{R}$. If we require that $\varphi^+(\cdot,t)$ be bounded as ${\rm Im}(x) \to +\infty$ for every $t \in \mathbb{R}$, then we must take $\lambda \geq 0$. The function $\varphi^+(\cdot,t)$ is $L$-periodic if and only if $\lambda \in \{ \lambda_n \}_{n = 1}^{\infty}$ with $\lambda_n := n$ for $n \in \mathbb{N}$. It has zero mean if and only if $\lambda \in \{ \mu_n \}_{n = 1}^{\infty}$ with $\mu_n := n+1$ for $n \in \mathbb{N}$. 
This proves (P1) and (P2) for $\{ \lambda_n \}_{n=1}^{\infty}$ and $\{ \mu_n \}_{n=1}^{\infty}$. \\

If both components $\varphi^+$ and $\varphi^-$ are nonzero, we
can separate the variables as 
$$
\varphi^{\pm}(x,t) = m^{\pm}(\xi) e^{-i\Omega t}.
$$ 
The system (\ref{Lax-pair}) is rewritten in ordinary derivatives:
\begin{equation}
\label{Lax-pair-ode}
\left\{ \begin{array}{l} i (m^+)' + \lambda( m^+ - m^-) + u m^+ = 0, \\
- (m^{\pm})''+ i (c + 2\lambda) (m^{\pm})' + [\pm i u' + H(u')] m^{\pm} = \Omega m^{\pm},
\end{array} \right.
\end{equation}
where $u = u(\xi)$ is given by (\ref{per-wave}). Since the second equation 
of the system (\ref{Lax-pair-ode}) is the second-order ODE, a general solution 
can be obtained as a superposition of two linearly independent solutions: 
\begin{equation*} 
m^+(\xi) = e^{i (\theta + \lambda + \frac{1}{2}c) \xi} \frac{(2\theta + k) + (2\theta-k) e^{i k \xi - \phi}}{1 + e^{ik \xi - \phi}} A_1 + 
e^{i (-\theta + \lambda + \frac{1}{2}c) \xi} \frac{(2\theta - k) + (2\theta+k) e^{i k \xi - \phi}}{1 + e^{ik \xi - \phi}} A_2
\end{equation*}
and 
\begin{equation*} 
m^-(\xi) = e^{i (\theta + \lambda + \frac{1}{2}c) \xi} \frac{(2\theta + k) + (2\theta-k) e^{i k \xi + \phi}}{1 + e^{ik \xi + \phi}} B_1 + 
e^{i (-\theta + \lambda + \frac{1}{2}c) \xi} \frac{(2\theta - k) + (2\theta+k) e^{i k \xi + \phi}}{1 + e^{ik \xi + \phi}} B_2,
\end{equation*}
where $\theta > 0$ is found from $\theta^2 = \Omega + (\lambda + \frac{1}{2} c)^2$ and the constants $A_{1,2}$, $B_{1,2}$ are arbitrary. Substituting the expressions for $m^{\pm}$ into the first equation of the system (\ref{Lax-pair-ode}) only defines equation between the constants:
\begin{equation*}
2 \lambda B_1 = -(2 \theta + c) A_1, \quad 2 \lambda B_2 = (2 \theta - c) A_2,
\end{equation*}
whereas $\lambda$ and $\theta$ are left arbitrary. 

If we require $\varphi^{\pm}(\cdot,t)$ to be bounded as ${\rm Im}(x) \to \pm \infty$, then we must set uniquely $\theta = |\lambda + \frac{1}{2} c|$ 
which yields $\Omega = 0$ and $\varphi^{\pm}(x,t) = m^{\pm}(\xi)$. For this choice of $\theta$, the linear superposition of two solutions yields only one solution in the form (after taking $A_{1,2} = -1$):
\begin{equation}
\label{m-plus} 
m^+(\xi) = \frac{(2\lambda + c - k) + (2\lambda + c + k) e^{i k \xi - \phi}}{1 + e^{ik \xi - \phi}}
\end{equation}
and 
\begin{equation}
\label{m-minus} 
m^-(\xi) = \frac{(2\lambda + c - k) + (2\lambda + c + k) e^{i k \xi + \phi}}{1 + e^{ik \xi + \phi}}.
\end{equation}
Parameter $\lambda$ is left arbitrary, whereas we can see that $\varphi^{\pm}(\cdot,t) = m^{\pm}(\cdot - c_0t)$ are $L$-periodic and $\mathbb{C}^{\pm}$-analytic for every $t \in \mathbb{R}$ and every $\lambda \in \mathbb{R}$. This proves (P3).

It remains to confirm (P1) and (P2) for $\lambda_0$ and $\mu_0$. If $\lambda = \lambda_0 = -(c+k)/2$, then the eigenfunction (up to a norming factor) is found from (\ref{m-plus}) and (\ref{m-minus}) as
\begin{equation}
\label{eigenvector-1}
m^+(\xi) = \frac{1}{1 + e^{ik \xi - \phi}}, \qquad 
m^-(\xi) = \frac{1}{1 + e^{ik \xi + \phi}} = \frac{e^{-ik\xi - \phi}}{1 + e^{-ik \xi - \phi}}.
\end{equation}
It follows from (\ref{eigenvector-1}) that the mean-value of $m^-$ is zero. If $\lambda = \mu_0 = -(c-k)/2$, then the eigenfunction (up to a norming factor) is found from (\ref{m-plus}) and (\ref{m-minus}) as
\begin{equation}
\label{eigenvector-2}
m^+(\xi) = \frac{e^{ik \xi - \phi}}{1 + e^{ik \xi - \phi}}, \qquad 
m^-(\xi) = \frac{e^{ik \xi + \phi}}{1 + e^{ik \xi + \phi}} = \frac{1}{1 + e^{-ik \xi - \phi}}.
\end{equation}
It follows from (\ref{eigenvector-2}) that the mean-value of $m^+$ is zero. 
For other values of $\lambda \in \mathbb{R}$, the mean values of $m^+$ and $m^-$ defined by (\ref{m-plus}) and (\ref{m-minus}) are nonzero. This proves (P1) and (P2) for $\lambda_0$ and $\mu_0$. Note that (\ref{m-plus}) and (\ref{m-minus}) agree with (1.11) and (1.22) in \cite{DK-91}.

\section{Derivation of the multi-breather solutions}
\label{sec-4}

\begin{figure}[htb!]
	\centering
	\includegraphics[width=12cm,height=6cm]{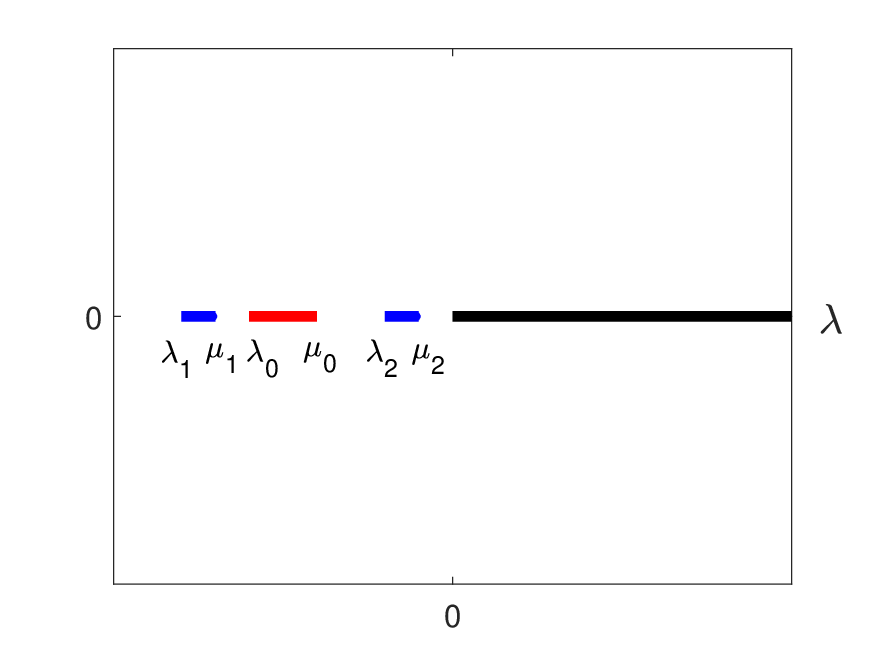}\\
	\includegraphics[width=12cm,height=6cm]{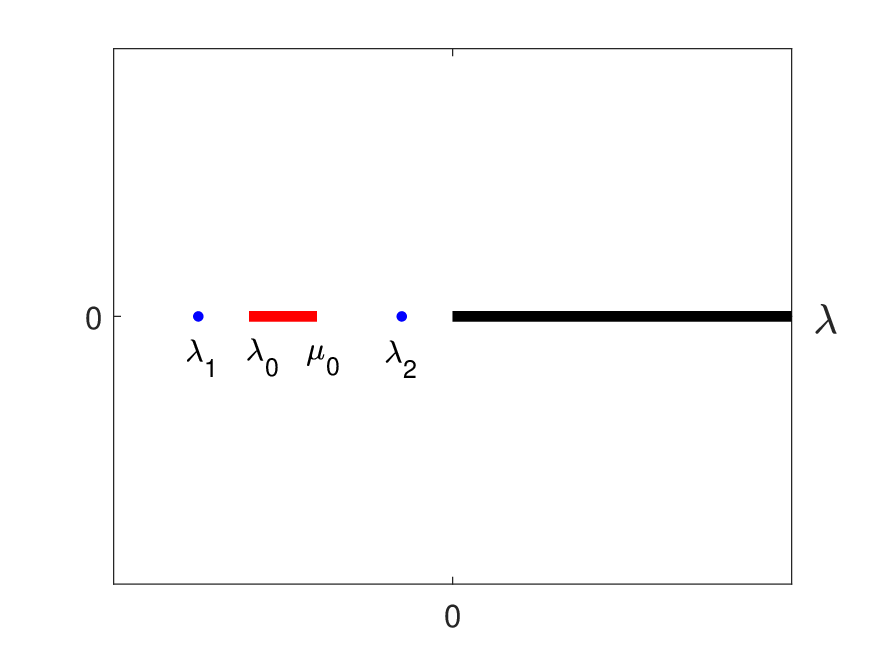}
	\caption{The degeneration procedure used to obtain multi-breather solutions. 
		Top: three finite bands in the Lax spectrum associated with the three-periodic solutions. 
		Bottom: two eigenvalues and one finite band in the Lax spectrum associated with the two-breather solution on the periodic wave. }
	\label{fig-bands}
\end{figure}

To obtain the multi-breather solutions of the BO equation 
(\ref{BO}) on the background of the traveling wave (\ref{per-wave}), we define solutions in the form (\ref{Hirota}) with $f = \det(F)$ and $f' = \det(F')$ 
being determinants of the square matrices $F$ and $F'$  
with the following elements:
\begin{align}
\label{sol-1}
F_{jm} &= \displaystyle \frac{1}{k_j} \exp\left( i k_j \xi_j + \phi_j + \frac{1}{2} \sum\limits_{m \neq j} A_{jm}\right) \delta_{jm} + \frac{2}{k_j + k_m + c_j - c_m}, \\
\label{sol-2}
F_{jm}' &= \displaystyle \frac{1}{k_j} \exp\left( i k_j \xi_j - \phi_j + \frac{1}{2} \sum\limits_{m \neq j} A_{jm} \right) \delta_{jm} + \frac{2}{k_j + k_m + c_j - c_m},
\end{align}
where $\xi_j = x - c_j t - x_j$, $k_j \phi_j > 0$, $x_j \in \mathbb{R}$, $c_j = k_j \coth \phi_j > 0$, and 
\begin{align}
\label{sol-3}
e^{A_{jm}} = \frac{(c_j-c_m)^2 - (k_j - k_m)^2}{(c_j-c_m)^2 - (k_j + k_m)^2}.
\end{align}
These determinant representations were derived in \cite[Eqs. (2.16)]{Matsuno04} based on the determinant formulas used in \cite{DK-91}. They are particularly useful in the derivation of multi-soliton solutions by degeneration of $k_j \to 0$ with $c_j$ being fixed, see \cite[Eqs. (2.48)]{Matsuno04}. It also follows from the representation, see \cite[Eq.(2.29), Appendix A]{Matsuno04},
\begin{equation}
\label{Matsuno-formula}
f' = \det(F') = \exp\left( \sum_{j} i k_j \xi_j - \phi_j \right) \det(\bar{F}) = \exp\left( \sum_{j} i k_j \xi_j - \phi_j \right) \bar{f}
\end{equation}
that the solution $u(x,t)$ in the form (\ref{Hirota}) is a real quantity given by 
$$
u = -\sum_{j} k_j + i \partial_x \log \frac{\det(\bar{F})}{\det(F)},
$$
where the bar denotes the complex conjugation. Moreover, it was shown in Lemma 1.1 of \cite{DK-91} that the solution $u(x,t)$ is bounded for every $(x,t) \in \mathbb{R}\times \mathbb{R}$ and the assumption on zeros of $f$ ($f'$) in the upper (lower) half-plane is satisfied if parameters of the solution satisfy the constraints 
\begin{align}
\label{sol-4}
(c_j - c_m)^2 > (|k_j| + |k_m|)^2, \quad j \neq m. 
\end{align}
Constraints (\ref{sol-4}) ensure that the Lax spectrum of the multiperiodic solutions is located on 
$$
\cup_{j} [\lambda_j,\mu_j]  \cup [0,\infty), \quad \lambda_j = -\frac{c_j+ |k_j|}{2}, \quad \mu_j = -\frac{c_j-|k_j|}{2}
$$
with the spectral band $[\lambda_j,\mu_j]$ being disjoint from the other spectral bands \cite{DK-91,GK-21}. See the upper panel of Figure \ref{fig-bands} for illustration. \\

{\bf Multi-breather solutions.} For multi-breather solutions, we consider the square matrices $F$ and $F'$ of size $(N+1)\times(N+1)$ and reserve the first element for the traveling periodic wave (\ref{per-wave}) with $\xi = x - c_0 t - \xi_0$, $k_0 = k > 0$, and $\phi_0 = \phi > 0$. For the remaining $N$ elements, we use enumeration $1 \leq j \leq N$, add the phase transformations $x_j \to x_j + \pi k_j^{-1}$, denote $\eta_j = x - c_j t - x_j$, and take the limit $k_j \to 0$, $\phi_j \to 0$ with $c_j = k_j \coth \phi_j > 0$ being fixed. This degeneration procedure transforms elements of $F$ and $F'$ to the form:
\begin{align*}
F_{00} &= \frac{1}{k} \left[1 + e^{i k \xi + \phi} \right], \quad F_{0m} = \frac{2}{k + c_0 - c_m}, \quad F_{m0} = \frac{2}{k + c_m - c_0},\\
F_{00}' &= \frac{1}{k} \left[ 1 + e^{i k \xi - \phi} \right], \quad F_{0m}' = \frac{2}{k + c_0 - c_m}, \quad F_{m0}' = \frac{2}{k + c_m - c_0},
\end{align*}
and for $1 \leq j \neq m \leq N$:
\begin{align*}
F_{jj} &= - i \eta_j - \frac{1}{c_j} - \frac{2k}{(c_0 - c_j)^2 - k^2},  \quad F_{jm} = \frac{2}{c_j - c_m}, \\
F_{jj}' &= - i \eta_j + \frac{1}{c_j} - \frac{2k}{(c_0 - c_j)^2 - k^2}, \quad F_{jm}' = \frac{2}{c_j - c_m}.
\end{align*}
The terms with $e^{A_{jm}}$ with $1 \leq j,m \leq N$ do not contribute to the limit because of the power expansion:
$$
e^{A_{jm}} = 1 + \frac{4 k_j k_m}{(c_j-c_m)^2} + \mathcal{O}(k_j^2 k_m^2).
$$
The terms with $e^{A_{0j}}$ with $1 \leq j \leq N$ do not contribute to the limit for $F_{00}$ but contribute to the limit for $F_{jj}$ because of the power expansion:
$$
e^{A_{0j}} = 1 + \frac{4 k k_j}{(c_0-c_j)^2 - k^2} + \mathcal{O}(k_j^2).
$$
The choice of $c_j$ for $1 \leq j \leq N$ is arbitrary as long as 
$|c_j - c_0| > k$ and all $c_j$ are distinct. These constraints follow from (\ref{sol-4}). After the degeneration, the Lax spectrum of the breather solutions consist on 
$$
\cup_{j=1}^N \lambda_j \cup [\lambda_0,\mu_0] \cup [0,\infty), \quad \lambda_0 = -\frac{c_0+k}{2}, \quad \mu_0 = -\frac{c_0-k}{2}, \quad \lambda_j = -\frac{c_j}{2}, 
$$
where simple eigenvalues $\{ \lambda_j \}_{j=1}^N$ are located either in $(-\infty,\lambda_0)$ or $(\mu_0,0)$. See the lower panel of Figure \ref{fig-bands} for illustration. The degeneration procedure has been used for the KdV equation in \cite{Bertola}. It has been widely used to obtain multi-soliton solutions from the multi-periodic solutions of the BO equation \cite{Matsuno04,Satsuma79}.

Multiplying the rows of $F$ and $F'$ by $k, c_1, \dots, c_N$ respectively yields the solution formula (\ref{Hirota-intro}) in Theorem \ref{theorem-main}. The solution $u(x,t)$ is real which follows from (\ref{Matsuno-formula}) in the same degeneration limit as 
\begin{equation}
f' = \det(F') = (-1)^N e^{i k \xi - \phi} \det(\bar{F}) = (-1)^N e^{i k \xi - \phi} \bar{f}.
\label{formula-der}
\end{equation}

Since (\ref{sol-1}), (\ref{sol-2}), (\ref{sol-3}), and (\ref{sol-4})  yield roots of $f = \det(F)$ ($f' = \det(F')$) in the upper (lower) half-plane for every $k_j > 0$ in Lemma 1.1 of \cite{DK-91}, the roots can only approach the real line but cannot cross the real line in the limit $k_j \to 0$. Since $f'$ is a related to the complex conjugation of $f$ by (\ref{formula-der}), the roots of $f$ and $f'$ on the real line coincide up to their multiplicities and the quotient in the solution form 
(\ref{Hirota}) can be used to redefine $f$ and $f'$ free of roots on the real line. This proves that the solution $u(x,t)$ arising in the limit $k_j \to 0$ is bounded for every $(x,t) \in \mathbb{R}\times \mathbb{R}$.

Similar to Lemma 1.1 in \cite{DK-91}, it should be possible to represent the solution of the Lax equations (\ref{Lax-pair}) for the multi-breather solutions of Theorem \ref{theorem-main} and prove rigorously that the individual determinants of the matrices $F$ and $F'$ given below (\ref{Hirota-intro}) do not vanish for every $(x,t) \in \mathbb{R}\times \mathbb{R}$. This would rule out the possibility of coalescence of roots of $f$ and $f'$ at the real axis in the limit $k_j \to 0$. The proof of the above fact is left out for further work. \\

{\bf Example with $N = 1$.} Expanding the determinants in Theorem \ref{theorem-main} for $N = 1$, we obtain 
\begin{align*}
\det(F) &= -(1+e^{ik\xi + \phi}) (i c\eta + 1 + \beta) + 2 \beta, \\
\det(F') &= -(1+e^{ik\xi - \phi}) (i c\eta - 1 + \beta) + 2 \beta, 
\end{align*}
where $\beta$ is given by (\ref{b-definition}) and 
$\eta = x - ct - \eta_0$ is defined with $c_1 = c$. 
These expressions for $f = \det(F)$ and $f' = \det(F')$ are equivalent to those in (\ref{expansion-for-1}) and lead to the exact solution (\ref{breather}) shown on Figure \ref{fig-sol-surface} for either $c > c_0 + k$ or $c < c_0 - k$. 
The explicit expressions for $\det(F)$ and $\det(F')$ agree with (\ref{formula-der}) with $N = 1$. \\

{\bf Example with $N = 2$.} Expanding the determinants in Theorem \ref{theorem-main} for $N = 2$ and simplifying the answers, we obtain 
\begin{align*}
\det(F) &=  (1+e^{ik\xi + \phi}) \left[ (1 + ic_1 \eta_1) (1 + i c_2 \eta_2) + \frac{4 c_1 c_2}{(c_1-c_2)^2} + \beta_1 \beta_2 \right] \\
& \quad +  (e^{ik\xi + \phi} - 1) \left[ \beta_1 (1 + i c_2 \eta_2) + \beta_2 (1 + i c_1 \eta_1) \right], \\
\det(F') &= (1+e^{ik\xi - \phi}) \left[ (1 - ic_1 \eta_1) (1 - i c_2 \eta_2) + \frac{4 c_1 c_2}{(c_1-c_2)^2} + \beta_1 \beta_2 \right] \\
& \quad + (1 - e^{ik\xi - \phi}) \left[ \beta_1 (1 - i c_2 \eta_2) + \beta_2 (1 - i c_1 \eta_1) \right],
\end{align*}
where 
$$
\beta_j := \frac{2k c_j}{(c_j-c_0)^2 - k^2}, \quad 1 \leq j \leq 2.
$$
The explicit expressions for $\det(F)$ and $\det(F')$ agree with (\ref{formula-der}) with $N = 2$. By using these formulas in (\ref{Hirota}), we obtain three different solutions for three choices of $c_1$ and $c_2$: a bright--bright breather if $c_1,c_2 > c_0 + k$, a bright-dark breather if $c_1 < c_0 - k$ and $c_2 > c_0 + k$, and a dark-dark breather if $c_1, c_2 < c_0 - k$. The solution surfaces for these breather solutions are shown on Figure \ref{fig-two-solitons}. \\

\vspace{0.25cm}

{\bf Acknowledgement.}  This work was supported in part by the National Natural Science Foundation of China (No. 12371248) and the Project “333” of Jiangsu Province. The authors thank A. Abanov for asking questions if the breathers similar to those in the KdV equation also exist in the BO equation. The authors also thank P. G\'{e}rard and P. Miller for discussions of the Lax spectrum related to the periodic wave solutions and Y. Matsuno for discussions of the 
zeros of $f$ and $f'$.


\begin{thebibliography}{99}
	
\bibitem{Abanov1} A. G. Abanov, E. Betterlheim, and P. B. Wiegmann, ``Integrable hydrodynamics of Calogero--Sutherland model: bidirectional Benjamin--Ono equation", J. Phys. A: Math. Theor. {\bf 42} (2009) 135201

\bibitem{Abanov} A. G. Abanov, A. Gromov, and M, Kulkarni, ``Soliton solutions of a Calogero model in a harmonic
potential", J. Phys. A: Math. Theor. {\bf 44} (2011) 295203 (21 pages)
	
\bibitem{Abanov2} A. G. Abanov and P. B. Wiegmann, ``Quantum hydrodynamics, the quantum Benjamin--Ono equation, and the Calogero model", Phys. Rev. Lett. 
{\bf 95} (2005) 076402

\bibitem{Cole} M. J. Ablowitz, J. T. Cole, G. A. El, M. A. Hoefer, and
X. Luo, ``Soliton-mean field interaction in Korteweg-de Vries
dispersive hydrodynamics,'' Stud. Appl. Math. {\bf 151} (2023) 795--856

\bibitem{Bertola} M. Bertola, R. Jenkins, and A. Tovbis, 
``Partial degeneration of finite gap solutions to the Korteweg--de
Vries equation: soliton gas and scattering on elliptic background", 
Nonlinearity {\bf 36} (2023) 3622--3660

\bibitem{BH14} J. C. Bronski and V. M. Hur, ``Modulational instability and variational structure", Stud. Appl. Math. {\bf 132} (2014) 285--331

\bibitem{Case79} K. M. Case, ``Meromorphic solutions of the Benjamin--Ono equation", Phys. A {\bf 96} (1979) 173--182

\bibitem{DK-91} S. Yu. Dobrokhotov and I. M. Krichever, ``Multiphase solutions of the Benjamin-Ono equation and their averaging", Math. Notes {\bf 49} (1991) 583--594

\bibitem{Smyth18} G. A. El, L. T. K. Ngueyn, and N. F. Smyth, 
``Dispersive shock waves in systems with nonlocal dispersion of 
Benjamin--Ono type", Nonlinearity {\bf 31} (2018) 1392--1416

\bibitem{AF83} A. S. Fokas and M. J. Ablowitz, ``The inverse 
scattering transform for the Benjamin--Ono equation: a pivot to 
multidimensional problems", Stud. Appl. Math. {\bf 68} (1983) 1--10

\bibitem{Trillo13} J. Garnier, G. Xu, S. Trillo, and A. Picozzi, 
``Incoherent dispersive shocks in the spectral evolution of 
random waves", Phys. Rev. Lett. {\bf 111} (2013) 113902 (5 pages)

\bibitem{GK-21} P. G\'{e}rard and T. Kappeler, ``On the integrability of the Benjamin--Ono equation on the torus", Comm. Pure Appl. Math. {\bf 74} (2021) 1685--1747

\bibitem{GKT-20} P. G\'{e}rard, T. Kappeler, and P. Topalov, ``On the spectrum of the Lax operator of the Benjamin--Ono equation on the torus", J.  Funct. Anal. {\bf 279} (2020) 108762 (75 pages)

\bibitem{GKT-22} P. G\'{e}rard, T. Kappeler, and P. Topalov, ``On the Benjamin--Ono equation on $\mathbb{T}$ and its periodic and quasiperiodic solutions", J. Spectr. Theory {\bf 12} (2022) 169--193

\bibitem{Gesztesy} 	F. Gesztesy and R. Svirsky, ``(m)KdV solitons on the background of quasi-periodic finite-gap solutions", Mem. Amer. Math. Soc. 
{\bf 118} (1995) 563 (88 pages)

\bibitem{Grimshaw} R. Grimshaw, ``Evolution equations for long nonlinear internal waves in stratified shear flows", Stud. Appl. Math. {\bf 65} (1981) 159--188

\bibitem{GSS21} R. H. J. Grimshaw, N. F. Smyth, and Y. A. Stepanyants, ``Interaction of internal solitary waves with long periodic waves within
the rotation modified Benjamin–Ono equation", Physica D {\bf 419} 
(2021) 132867 (10 pages)



\bibitem{HMP} M. Hoefer, A. Mucalica, and D. E. Pelinovsky, ``KdV breathers on cnoidal wave background", J. Phys. A: Mathem. Theor. {\bf 56} (2023) 185701 (25 pages)



\bibitem{Smyth99} M. C. Jorge, A. A. Minzoni, and N. F. Smyth, ``Modulation solutions for the Benjamin–Ono equation", Physica D {\bf 132} (1999) 1--18


\bibitem{KM98} D. J. Kaup and Y. Matsuno, ```The inverse 
scattering transform for the Benjamin--Ono equation", Stud. Appl. Math. {\bf 101} (1998) 73--98

\bibitem{Kuznetsov} E. A. Kuznetsov and A. V. Mikhailov, ``Stability of stationary waves in nonlinear weakly dispersive
media", Sov. Phys. JETP {\bf 40} (1974)  855--859

\bibitem{Ling} L. Ling and X. Sun, ``Multi-elliptic-dark soliton solutions in the defocusing nonlinear Schr\"{o}dinger equation", Appl. Math. Lett. {\bf 148} (2023) 108866 (9 pages)

\bibitem{Lou} S. Lou, X. Cheng, and X. Tang, ``Dressed dark solitons 
of the defocusing nonlinear Schr\"{o}dinger equation", Chinese Phys. Lett. {\bf 31} (2014) 070201 (4 pages)

\bibitem{Maiden}  M. D. Maiden, D. V. Anderson, A. A. Franco,
G. A. El, and M. A. Hoefer, ``Solitonic dispersive hydrodynamics:
Theory and observation,'' Phys. Rev. Lett. {\bf 120} (2018) 144101 

\bibitem{Matsuno82} Y. Matsuno, ``Exact multi-soliton solution of the Benjamin--Ono equation", J. Phys. A : Math. Gen. {\bf 12} (1979) 619--621

\bibitem{Matsuno-book} Y. Matsuno, {\em Bilinear Transformation Method} 
(Academic Press, New York, 1984) 

\bibitem{Matsuno98} Y. Matsuno, ``Nonlinear modulation of periodic waves in the small dispersion limit of the Benjamin--Ono equation", Phys. Rev. E {\bf 58} (1998) 7934--7940

\bibitem{MATSUNO} Y. Matsuno, ``The small dispersion limit of the Benjamin--Ono equation and the evolution of a step initial condition", J. Phys. Soc. Japan {\bf 67} (1998) 1814--1817

\bibitem{Matsuno04} Y. Matsuno, ``New representations of multiperiodic and multisoliton solutions for a class of nonlocal soliton equations", J. Phys. Soc. Japan {\bf 73} (2004) 3285--3293

\bibitem{MSKK} Y. Matsuno, V. S. Shchenovich, A. M. Kamchatnov and R.A. Kraenkel, ``Whitham method for the Benjamin--Ono--Burgers equation and dispersive shocks", Phys. Rev. E {\bf 75} (2007) 016307 (5 pages)

\bibitem{Mao} Y. Mao, S. Chandramouli, W. Xu, and M. Hoefer, ``Observation of traveling breathers and their scattering in a two-fluid system", Phys. Rev. Lett. {\bf 131} (2023) 147201 (7 pages)

\bibitem{Miloh} T. Miloh, ``On periodic and solitary wavelike solutions 
of the intermediate long-wave equation", J. Fluid Mech. {\bf 211} (1990) 617--627

%\bibitem{MP22} A. Mucalica and D.E. Pelinovsky, ``Solitons on the rarefaction wave background via the Darboux transformation", Proc. R. Soc. A {\bf 478} (2022) 20220474 (17 pages) 

\bibitem{HMP-24} A. Mucalica and D. E. Pelinovsky, ``Dark breathers on the snoidal wave background in the dofucusing mKdV equation", arXiv: 2312.08969 (2023)

\bibitem{Munoz1} C. Munoz and G. Ponce, ``Breathers and the dynamics of 
solutions in KdV type equations", Comm. Math. Phys. {\bf 367} (2019) 581--598

\bibitem{Munoz2} C. Munoz and G. Ponce, ``On the asymptotic behavior of solutions to the Benjamin--Ono equation", Proceedings of the AMS {\bf 147} (2019) 5303--5312

\bibitem{Nakamura79} A. Nakamura, ``B\"{a}cklund transform and conservation laws of the Benjamin--Ono equation", J. Phys. Soc. Japan {\bf 47} (1979) 1335--1340

\bibitem{Smyth21} L. T. K. Ngueyn and N. F. Smyth, 
``Dispersive shock waves for the Boussinesq Benjamin–Ono equation", 
Stud. Appl. Math. {\bf 147} (2021) 32--59

\bibitem{P95} D. Pelinovsky, ``Intermediate nonlinear Schro\"{o}dinger equation for internal waves in a fluid of finite depth", Phys. Lett. A {\bf 197} (1995) 401--406

\bibitem{Smyth02} A. Porter and N. F. Smyth, ``Modelling 
the morning glory of the Gulf of Carpentaria", J. Fluid Mech. {\bf 454} (2002) 1--20

\bibitem{Satsuma79} J. Satsuma and Y. Ishimori, ``Periodic wave and rational soliton solutions of the Benjamin--Ono equation", J. Phys. Soc. Japan {\bf 46} (1979) 681--687

\bibitem{Shin} H. Shin, ``The dark soliton on a cnoidal wave background", 
J. Phys. A: Math. Gen. {\bf 38} (2005) 3307--3315

\bibitem{MS94} M. D. Spector and T. Miloh, ``Stability of nonlinear periodic internal waves in a deep stratified fluid", SIAM J. Appl. Math. 
{\bf 54} (1994) 688--707

\bibitem{Sprenger} P. Sprenger, M. A. Hoefer, and G. A. El, ``Hydrodynamic optical soliton tunneling", Phys. Rev. E {\bf 97} (2018)  032218

\bibitem{Takahashi} D. A. Takahashi, ``Integrable model for density-modulated quantum condensates: Solitons passing through a soliton lattice``, Phys. Rev E. 93 (2016) 062224 (20 pages)
		
\end{thebibliography}
\end{document}